\documentclass[conference]{IEEEtran}
\IEEEoverridecommandlockouts

\usepackage{cite}
\usepackage{amsmath,amssymb,amsfonts}
\usepackage{algorithmic}
\usepackage{graphicx}
\usepackage{textcomp}
\usepackage{xcolor}
\usepackage{subcaption}
\usepackage{booktabs}

\def\BibTeX{{\rm B\kern-.05em{\sc i\kern-.025em b}\kern-.08em
    T\kern-.1667em\lower.7ex\hbox{E}\kern-.125emX}}
\begin{document}

\title{SYKI-SVC: Advancing Singing Voice Conversion with Post-Processing Innovations and an Open-Source Professional Testset\\

\thanks{$^{\dag}$Corresponding author.}
}

\author{\IEEEauthorblockN{Yiquan Zhou}
\IEEEauthorblockA{\textit{School of Software Engineering} \\
\textit{Xi’an Jiaotong University}\\
Xi'an, China \\
yiqian.zhou@stu.xjtu.edu.cn}
\and
\IEEEauthorblockN{Wenyu Wang}
\IEEEauthorblockA{\textit{School of Software Engineering} \\
\textit{Xi’an Jiaotong University}\\
Xi'an, China \\
wenyu.wang@stu.xjtu.edu.cn}
\and
\IEEEauthorblockN{Hongwu Ding}
\IEEEauthorblockA{\textit{AI Center Speech Group} \\
\textit{Happy Elements}\\
Shanghai, China \\
E22201087@stu.ahu.edu.cn}
\and
\IEEEauthorblockN{Jiacheng Xu}
\IEEEauthorblockA{\textit{School of Software Engineering} \\
\textit{East China Normal University}\\
Shanghai, China \\
xujiacheng28@outlook.com}
\and
\IEEEauthorblockN{Jihua Zhu$^{\dag}$}
\IEEEauthorblockA{\textit{School of Software Engineering} \\
\textit{Xi’an Jiaotong University}\\
Xi'an, China \\
zhujh@xjtu.edu.cn}
\and
\IEEEauthorblockN{Xin Gao}
\IEEEauthorblockA{\textit{Division of Music and Audio} \\
\textit{Union Wheatland Culture and Media Ltd.}\\
Chengdu, China \\
sui@musinya.com}
\and
\IEEEauthorblockN{Shihao Li}
\IEEEauthorblockA{\textit{Division of Music and Audio} \\
\textit{Union Wheatland Culture and Media Ltd.}\\
Chengdu, China \\
yiseho@yiseho.com}
}

\maketitle

\begin{abstract}
Singing voice conversion aims to transform a source singing voice into that of a target singer while preserving the original lyrics, melody, and various vocal techniques. In this paper, we propose a high-fidelity singing voice conversion system. Our system builds upon the SVCC T02 framework and consists of three key components: a feature extractor, a voice converter, and a post-processor. The feature extractor utilizes the ContentVec and Whisper models to derive F0 contours and extract speaker-independent linguistic features from the input singing voice. The voice converter then integrates the extracted timbre, F0, and linguistic content to synthesize the target speaker's waveform. The post-processor augments high-frequency information directly from the source through simple and effective signal processing to enhance audio quality. Due to the lack of a standardized professional dataset for evaluating expressive singing conversion systems, we have created and made publicly available a specialized test set. Comparative evaluations demonstrate that our system achieves a remarkably high level of naturalness, and further analysis confirms the efficacy of our proposed system design.

\end{abstract}


\section{Introduction}
Singing voice conversion (SVC) aims to transform a source singing voice into that of a target singer while preserving the original lyrics, melody, and various vocal techniques. This task is more challenging than general voice conversion (VC) due to the greater expressiveness inherent in singing~\cite{huang2023singing}.\footnote{This papper talks about svc as a many-to-one svc system}

A key challenge in SVC is the disentanglement and recombination of the speaker's timbre with the fundamental content and melody of the singing voice. Unlike the VC task, SVC requires the preservation of the singing techniques of the source voice as much as possible. To achieve this, several studies~\cite{kaneko2020cycleganvc3,kaneko2021MaskCycleGAN-VC,kameoka2018starganvc,kaneko2019starganvc2} have explored the use of Generative Adversarial Networks (GANs) to enhance SVC performance. However, these systems often face a trade-off between naturalness and speaker similarity, because the timbre and linguistic content in the voice are not sufficiently separated. To improve SVC performance, researchers have focused on adopting a recognition-synthesis paradigm. This approach leverages pre-trained models to extract speaker-independent features, which are then transformed into singing voices by the SVC model. Recent methods~\cite{ning2023vits,soft-vc-2022,Tian2020TheN,hwang2022stylevc} frequently utilize Bottleneck Features (BNFs) derived from Automatic Speech Recognition (ASR) models as speaker-independent intermediate representations. Meanwhile, self-supervised learning (SSL) models~\cite{hsu2021hubert,jayashankar2023self}, which are trained on extensive unlabeled speech datasets, have shown strong performance in next-generation speech recognition by extracting robust linguistic features. Research has started to integrate SSL features into SVC tasks, resulting in promising outcomes regarding naturalness and similarity. Some research~\cite{ning2023vits,zhou2023vits} has explored end-to-end frameworks that directly synthesize the converted singing voice waveforms to enhance fidelity; however, challenges persist in accurately reconstructing high-frequency details.

\begin{figure*}[t]
  \centering
  \begin{subfigure}[b]{.45\linewidth}
    \centering
    \includegraphics[width=\textwidth]{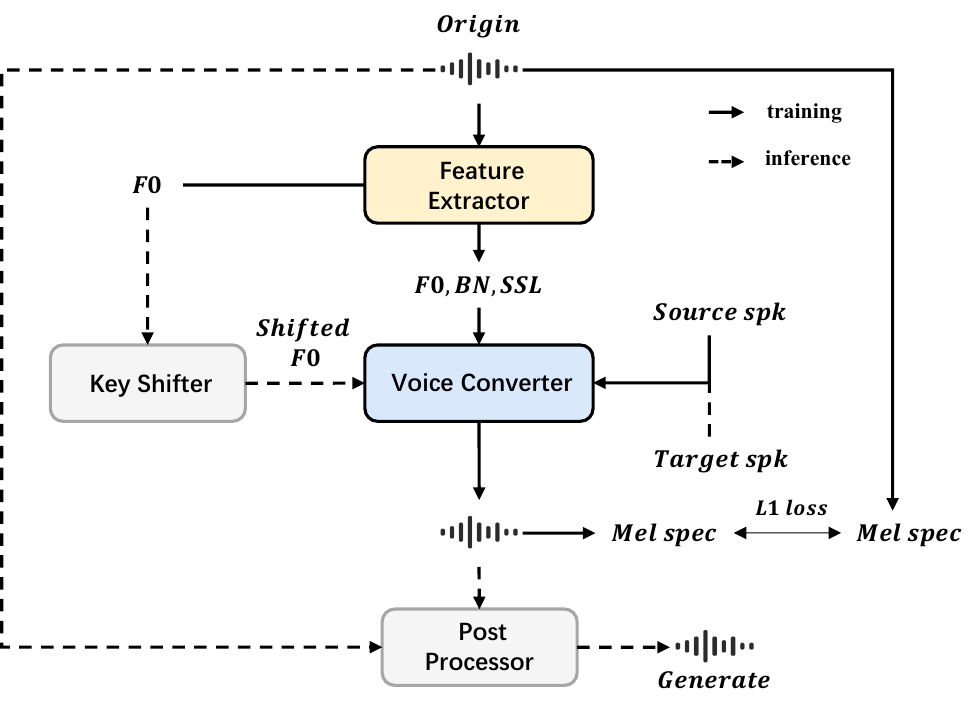}
    \caption{Overall}
    \label{fig:1(a)}
  \end{subfigure}
  \hfill
  \begin{subfigure}[b]{.45\linewidth}
    \centering
    \includegraphics[width=\textwidth]{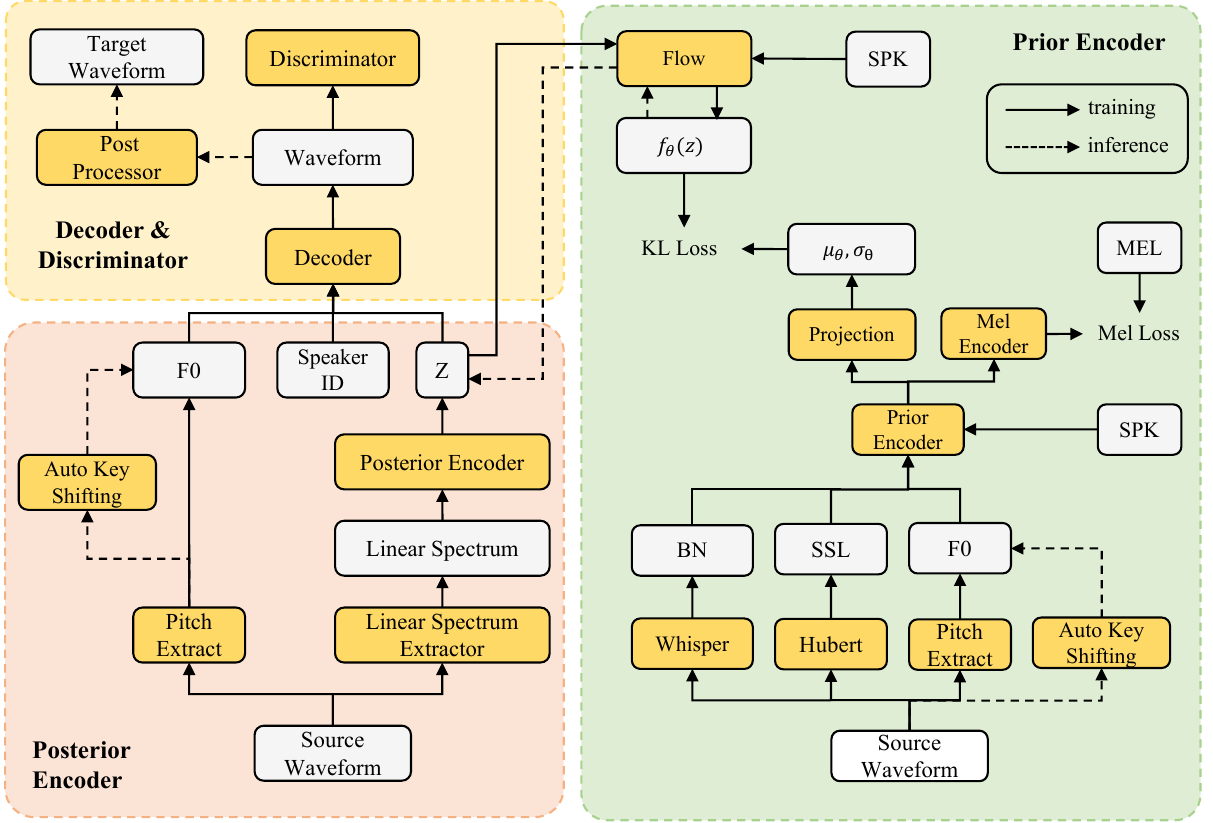}
    \caption{Model Details}
    \label{fig:1(b)}
  \end{subfigure}
  \caption{Fig. (a) shows the general description of the training inference, and Fig. (b) shows the details of the model}
  \label{fig:1}
\end{figure*}

In this paper, we introduce our SVC system, named SYKI-SVC. Our proposed model enhances the top-performing system from SVCC2023~\cite{huang2023singing}. Building upon the baseline system and adopting a recognition-synthesis approach, we utilize the SSL model ContentVec and the ASR model Whisper~\cite{radford2023robust} to extract speaker-independent features that represent the linguistic content of the source singing voice. Concurrently, the fundamental frequency (F0) and speaker ID capture the prosody of the voice and the timbre of the target singer, respectively. Our SVC model integrates these features to generate the target singing voice. Additionally, we synthesize the spectrum of the target singing voice using intermediate variables combined with the speaker ID to ensure that the intermediate variables retain as much information as possible. To further enhance our model’s performance in high-frequency regions, we introduce post-processing techniques. Our experiments demonstrate that the timbre information above 10 kHz is minimal; therefore, we directly supplement the high-frequency components of our synthesized singing voice with those from the source, thereby improving the overall quality of the generated voice.

Unlike VC tasks, singing requires unique techniques that only trained singers can master. To make SVC systems suitable for professional use, their ability to retain these techniques must be evaluated. Currently, there is a lack of a professional open-source test dataset. To address this, we released a dataset featuring various singing techniques performed by professionals. This allows researchers to better assess if their SVC models can preserve these techniques accurately.

Comparative experiments demonstrate that our system significantly enhances naturalness while maintaining a timbre similarity comparable to mainstream frameworks. Notably, after post-processing, is capable of generating singing audio that is consistent with the source audio, with a quality that can reach up to 48kHz. Samples of the converted voices from our system are available for listening on our platform.\footnote{https://ryker0923.github.io/sykisvc/}

\section{System Overview}

We made further enhancements based on the SVCC2023 T02 system~\cite{zhou2023vits}. Fig. \ref{fig:1(a)} illustrates our SVC system, which comprises two components: training and inference.

\textbf{Training}
The training phase includes two modules: the feature extractor and the singing voice converter. During training, the feature extractor extracts SSL features, BN features, F0, and linear spectrogram features from the singing voice. The singing voice conversion model then integrates these features as input and reconstructs the voice using the speaker ID.

\textbf{Inference}
The inference phase comprises three modules: the feature extractor, the singing voice converter, and the post-processing module. The singing voice converter utilizes the extracted SSL features, BN features, and modified F0 extracted in the feature extractor as inputs to generate the target singing voice with the desired speaker's timbre. To further enhance the quality of the generated voice, the post-processor directly incorporates the high-frequency components from the source singing voice into the generated target singing voice.

\section{Proposed Methods}
The overall architecture of SYKI-SVC is shown in Fig. \ref{fig:1(b)}. It consists of three main parts: a feature extractor, a singing voice converter, and a post-processor. Each of these components will be described in detail in the following sections.
\subsection{Feature Extractor}
In the Feature extractor, we used features from the SSL model ContentVec~\cite{qian2022contentvec} and the BNF from the ASR model Whisper~\cite{radford2023robust}. Both are utilized to separate the speaker's timbre from the linguistic content of the singing audio. The features from ContentVec are fused with Whisper using element-wise addition as the input for SYKI-SVC. For speaker identity, we use a look-up table to learn speaker embedding. In addition, SKYI-SVC employs a neural network-based fundamental frequency (F0) extractor, RMVPE~\cite{wei2023rmvpe}, to calculate the F0 for generating the accurate melody of the converted singing voice.

\subsection{Singing Voice Converter}

We implemented the SVC model based on VITS~\cite{kim2021conditional}, comprising a posterior encoder, a prior encoder, a decoder, and a discriminator. During training, the posterior encoder converts the source waveform $ y $ into a latent representation $ z $, modeling the posterior distribution $ P(z|y) $. The decoder then reconstructs $ z $ into the original waveform, forming a self-reconstruction process. Our study adheres to the baseline structure, enhancing singing voice quality by integrating an F0-based sinusoidal signal into the HiFi-GAN~\cite{kong2020hifigan} decoder. The prior encoder captures features like timbre and pitch, transforming them via normalizing flows into the posterior distribution. For F0 features, a difference of 4 keys is applied when converting between singers of different genders. Inspired by~\cite{zhang2022visinger}. To prevent information loss in the posterior encoder, we added a module introducing an additional supervisory loss, as detailed in Equations \ref{1}, \ref{2}, and \ref{3}. This module uses the latent representation $ z $ and speaker ID to reconstruct the Mel spectrogram. Better reconstruction of the Mel spectrogram indicates that the latent variable $ z $ retains more content information. During inference, the model achieves singing voice conversion using the prior encoder and decoder to apply the target timbre.
\begin{equation}
	mel_p = pre(z)\label{1} \\
\end{equation}
\begin{equation}
	loss_{mel} = |mel_p - mel|_{l1}\label{2} \\
\end{equation}
\begin{equation}
	loss_{all} = loss_{remain} + loss_{mel}\label{3}\\
\end{equation}

 $loss_{remain}$ represents all losses in the baseline model.

\subsection{Post-Processor}

Synthesizing high-frequency information has always posed challenges for SVC and VC tasks. However, predicting these frequencies through the network may not be necessary. We found that the portions of the waveform with frequencies above 10 kHz typically contain minimal timbral information. Since these high-frequency bands do not influence the timbre, and the frame levels of the source and target songs are aligned, we can directly supplement the frequencies above 10 kHz from the source song into the output generated by the converter. This approach enhances the output quality without compromising its similarity. In this context, $ F_h $ represents the high-pass filter, and $ F_l $ denotes the low-pass filter. Let $ wav_s $ be the source speech, and $ wav_c $ be the converted speech. We supplement the high-frequency information using the following equation:
\begin{equation}
	Diff = mean(abs(wav_{s}))/ mean(abs(wav_{c}))\label{4}\\
\end{equation}
\begin{equation}
	Out = F_{h}(wav_{s})+  F_{l}(wav_{c})* Diff   \label{5}\\
\end{equation}

\section{Experimental}

\subsection{Experimental Design}
\begin{table*}
  \caption{The subjective evaluation results, expressed as MOS with 95\% confidence intervals, cover vocal naturalness, bite and technique reproduction, and tone similarity, while objective evaluations include Cosine Similarity.}
  \label{tab:1}
  \renewcommand{\arraystretch}{1.0}
  \begin{tabular}
   {l@{\hspace{15pt}}|c@{\hspace{25pt}}c@{\hspace{25pt}}c@{\hspace{25pt}}c@{\hspace{25pt}}|c@{\hspace{15pt}}}
    \midrule
    Approach & Vocal naturalness $\uparrow$ & Bite reproduction
 $\uparrow$ & Technique reproduction $\uparrow$ & Tone similarity
 $\uparrow$ & Cos.Sim $\uparrow$ \\
    \midrule
     DDSP-SVC & $2.42 \pm 0.09$ & $2.51 \pm 0.11$ & $2.42 \pm 0.11$ & $2.89 \pm 0.09$ & $0.8551$ \\
     so-vits & $3.67 \pm 0.09$ & $3.78 \pm 0.11$ & $3.85 \pm 0.11$ & $4.03 \pm 0.12$ & $0.8216$ \\
     SVCC2023-T23 & $3.92 \pm 0.09$ & $3.53 \pm 0.09$ & $3.85 \pm 0.08$ & $3.78 \pm 0.09$ & $0.8530$ \\
     SYKI-SVC& $\mathbf{4.11} \pm \mathbf{0.08}$ & $\mathbf{4.11} \pm \mathbf{0.13}$ & $\mathbf{4.09} \pm \mathbf{0.09}$ & $\mathbf{4.25} \pm \mathbf{0.11}$ & $0.8560$ \\
     \midrule
      - Mel-loss & $4.01 \pm 0.11$ & $3.96 \pm 0.13$ & $4.07 \pm 0.08$ & $4.17 \pm 0.09$ & $0.8526$ \\
      - feature fusion & $4.03 \pm 0.09$ & $3.53 \pm 0.08$ & $3.67 \pm 0.09$ & $4.21 \pm 0.07$ & $\mathbf{0.8602}$ \\
      - Post-processer & $3.82 \pm 0.11$ & $3.71 \pm 0.12$ & $3.82 \pm 0.08$ & $4.14 \pm 0.09$ & $0.8574$ \\
    \midrule
  \end{tabular}
\end{table*}

\begin{table}
    \caption{TestSet distribution}
    \label{tab:2}
    \renewcommand{\arraystretch}{1.0}
    \begin{tabular}{l@{\hspace{20pt}}l@{\hspace{20pt}}l@{\hspace{20pt}}l@{\hspace{20pt}}}
    \hline Technique & Duration(min) & Gender & Number\\
    \hline Vibrato  & 2.75 & FM &  13\\
    Breathy Voice & 2.2 & FM  & 13\\
    Growling  & 0.70 & FM & 4\\
    Crying Tone & 1.9 & FM & 9\\
    Sobbing  & 0.52& F & 3\\
    Falsetto  & 1.58 & FM & 9\\
    Strong Mix & 1.23 & FM & 6\\
    Weak Mix &1.33 & FM & 6\\
    Ultra High Pitch & 0.12 & F & 1\\
    Ultra Low Pitch & 0.67 & M & 3\\
    Pharyngeal Sound & 0.15 & F & 1\\
    Bubble Sound & 1.43 & FM & 8\\
    Portamento  & 2.82 & FM & 15\\
    Breathing Sounds  & 1.35 & M & 6\\
    Humming   & 2.40 & FM & 11\\
    Bel Canto  & 1:14 & FM & 7\\
    Folk  & 1.03 & F & 5\\
    Rap  & 0.7 & FM &2\\
    Rock  & 0.18 & F & 1\\
    Opera  & 0.98 & FM & 4\\
    Jazz  & 0.83 & F & 5\\
    \hline
    \end{tabular}
\end{table}

For songs that require more complexity and advanced skills, the test dataset is often limited, a randomly selected test set may not be a good validation of the model's upper bound. To address this limitation, we invited 6 high-level singers to record test data in a professional studio, incorporating a variety of techniques such as strong mixing, weak mixing, falsetto, and rap, among others. Moreover, we will make this test collection open source to provide valuable data for future research\footnote{https://pan.baidu.com/s/15Brj27-lDp2n9TvhBPvDQw?pwd=w9jc}\footnote{https://drive.google.com/file/d/1-DPhpSi9gcTFczKwtGLHed60S4XnhB3N/\\view?usp=drivel
ink}. The detailed information is recorded in Table \ref{tab:2}.

For the listening tests, we typically involve ordinary individuals to evaluate the songs. However, in this study, we aim to preserve the skills of the source during the conversion process. Non-musicians may lack the sensitivity to discern whether these skills are retained. Therefore, we selected 10 ordinary listeners and 10 professional musicians to participate in the listening experiment, ensuring that the results are both reasonable and reliable. 

We established four dimensions to evaluate the outcomes generated by the SVC system: vocal naturalness, bite reproduction, technique reproduction, and timbre similarity.

\subsubsection{Vocal naturalness} refers to whether a synthesized human voice sounds more like a human voice. Factors such as electronic or metallic sounds can influence this aspect. The listener needs to assess the naturalness on a five-point scale.

\subsubsection{Bite reproduction} refers to whether the synthesized vocal accurately restores the bite habits and audibility of the original source. During this process, the listener is presented with both the original audio and the converted audio. The listener then indicates whether the bite habits and audibility of the two samples are equivalent using a five-point scale.

\subsubsection{Technique reproduction} refers to whether the synthesized voice accurately reproduces the singing technique of the original source. During this process, the listener will hear both the original audio and the converted audio, and using a five-point scale, whether the technique of the converted audio effectively mirrors that of the original source.

\subsubsection{Tone similarity} refers to whether the timbre of the converted song is similar to the timbre of the target speaker's song. During this assessment, the listener compares the natural voice of the target speaker with the converted voice. The listener can determine on a five-point scale, whether the two samples are produced by the same speaker. At the same time we also evaluated the cosine similarity of the tones.

All the data provided to the test set were converted into target male and female voices using various models. The same test voice, processed through different models, was presented to the listeners in a blind test to conduct the experiment. At the conclusion of the experiment, the average score from all listeners was calculated and used as the final result.

\subsection{Data and Model Training}
For the feature extractor, we use the open-source model 256-dimensional ContentVec and 1024-dimensional Whisper meduim version. For the speech converter, in the pre-training phase we use VCTK~\cite{yamagishi2019vctk}, NUS48e~\cite{6694316}, Opencpop~\cite{wang2022opencpop} , M4singer~\cite{zhang2022msinger} and Opensinger~\cite{wang2022opencpop} as well as one of our closed-source databases of male voices benchmarked against opencpop. In the adaptation phase, we trained the volume of two speakers, one male and one female, for the female speaker, we directly used the opencpop dataset, while for the male speaker, we recorded a closed-source dataset. We used both datasets for fine-tuning and subsequent experiments. all resampled to 24k for training.
\subsection{Experimental Setting}
We extracted ContentVec and Whisper meduim features through 1-dimensional convolutional compression to 197 dimensions after the use of element summing as the fusion feature input. As for the transformation model, the posterior encoder uses a WaveNet residual block, the same as WaveGlow~\cite{prenger2018waveglow}. The a prior encoder is realized by a multilayer transformer ~\cite{vaswani2017attention}. The decoder follows the original configuration of the HIFI-GAN decoder in VITS. Meanwhile, the structure of the new predicted mel spectra added to SKYI-SVC is identical to that of the a posterior encoder. During the training process, we trained the conversion model on the hybrid speech dataset and the singing dataset with 600k and 300k steps, respectively, with a batch size of 72. The number of training steps in the adaptation phase was 100k. The conversion model was optimized using the Adam optimizer with an initial learning rate of 1e-4 . The entire model was trained on an NVIDIA A800 GPU. For the spkembedding objective metrics, we use the pre-trained speaker encoder from PPGVC~\cite{Liu2021}.

\subsection{Experimental Results}

We conducted both subjective and objective evaluations of the model. We compared DDSP-SVC\footnote{https://github.com/yxlllc/DDSP-SVC}, so-vits\footnote{https://github.com/svc-develop-team/so-vits-svc}, and the SVCC2023 T23-system \cite{ning2023vits}.We conducted the experimental testing utilizing the 4 evaluation metrics described above, in conjunction with the cosine similarity (Cos.Sim) of speaker embeddings, to accomplish the assessment. All experimental results are recorded in Table \ref{tab:1}.

\textbf{Comparison Experiment}
From both subjective and objective experimental results, our model achieved the highest scores in the dimensions of vocal naturalness, bite reproduction, technical fidelity, and timbre similarity, surpassing the performance of popular conversion models. It is noteworthy that our model was trained exclusively on data with a 24 kHz sampling rate and was subsequently upsampled to 48 kHz using post-processing techniques, without any additional training. Therefore, the results at 48 kHz in our comparative experiments are valid for comparison with other systems.

\textbf{Ablation Experiment}
To further validate the effectiveness of our model, we conducted several deletion experiments. Specifically, we removed the fusion features, mel spectral supervision loss, and post-processor in succession. When feature fusion was eliminated, timbre similarity remained essentially unchanged, while the other metrics exhibited a decline. This clearly demonstrates the necessity of integrating multiple features. The removal of mel signal supervision resulted in a slight decrease in bite reproduction. Additionally, eliminating the post-processing component led to a deterioration in sound quality, which subsequently degraded various metrics. These results indicate that each component is crucial to the overall performance of the system.

\section{Conclusion}
In this paper, we propose several optimization strategies for Source-Vocoder Conversion (SVC) to achieve high-fidelity sound quality. To improve articulatory accuracy, we introduce feature fusion and an additional supervised loss to minimize occlusion. Additionally, we design a post-processor that enhances the timbre of synthesized audio by supplementing missing source information. Experiments confirm the effectiveness of our methods. We also propose a comprehensive evaluation framework and a publicly available test set to facilitate further SVC research. Our system demonstrates state-of-the-art performance on this test set compared to popular systems.

\section{Acknowledgement}
This work was supported by the Key Research and Development Programof Shaanxi under Grant No. 2024GX-YBXM-556 and the HPC Platform of Xi’an Jiaotong University.

\bibliographystyle{IEEEtran}
\bibliography{mybib}
\end{document}